\documentclass[twocolumn, prl]{revtex4}

\usepackage{graphicx}
\usepackage{dcolumn}
\usepackage{bm}
\usepackage{color}

\newcommand{\correct}[1]{\textcolor{black}{#1}}

\begin{document}

\preprint{APS/123-QED}

\title{
 Quasiparticle Excitations  and Evidence for
  Superconducting Double  Transitions  in Monocrystalline 
  U$_{0.97}$Th$_{0.03}$Be$_{13}$
    \color{black} 
 }



\author{Yusei Shimizu}
\email{yuseishimizu@imr.tohoku.ac.jp}
 \altaffiliation[Present address:\ ]{Institute for Materials Research, Tohoku University, Oarai, Ibaraki,  311-1313, Japan.}
\affiliation{Institute for Solid State Physics, University of Tokyo, Kashiwa, Chiba 277-8581, Japan.}
\author{Shunichiro Kittaka}
\affiliation{Institute for Solid State Physics, University of Tokyo, Kashiwa, Chiba 277-8581, Japan.}
\author{Shota Nakamura}
\affiliation{Institute for Solid State Physics, University of Tokyo, Kashiwa, Chiba 277-8581, Japan.}
\author{Toshiro Sakakibara} 
\affiliation{Institute for Solid State Physics, University of Tokyo, Kashiwa, Chiba 277-8581, Japan.}
\author{ Dai Aoki } 
\affiliation{ Institute for Materials Research (IMR), Tohoku University, Oarai, Ibaraki  311-1313, Japan.}
\author{ Yoshiya Homma } 
\affiliation{ Institute for Materials Research (IMR), Tohoku University, Oarai, Ibaraki  311-1313, Japan.}
\author{ Ai Nakamura } 
\affiliation{ Institute for Materials Research (IMR), Tohoku University, Oarai, Ibaraki  311-1313, Japan.}
\author{  Kazushige Machida} 
\affiliation{Department of Physics, Ritsumeikan University, Kusatsu, Shiga 525-8577, Japan.}


\date{\today}%

\begin{abstract}
Superconducting (SC) gap symmetry  and magnetic response  of   cubic U$_{0.97}$Th$_{0.03}$Be$_{13}$ 
    \correct{are} studied by means of high-\correct{precision} 
 heat-capacity and dc magnetization measurements  using  a single crystal,
 in order to address the long-standing question \correct{of its second}  phase transition \correct{at $T_{ \rm c2}$}
  in the  SC  state below $T_{\rm c1}$.
The absence (presence) of an anomaly at $T_{\rm c2}$  in the field-cooling (zero-field-cooling) magnetization indicates that this  transition is between two different SC states.
There is a qualitative difference in the field variation of the transition temperatures; $T_{\rm c2}(H)$ is isotropic whereas $T_{\rm c1}(H)$ exhibits a weak anisotropy between [001] and [111] directions.
\correct{In the low temperature phase below $T_{\rm c2}(H)$, the angle-resolved}  heat-capacity $C(T, H, \phi)$ reveals that the gap is fully opened
 over the Fermi surface, narrowing down the possible gap symmetry.
\end{abstract}

\maketitle


The nature of superconductivity in heavy-fermion compounds is of primary importance because  
 an unconventional pairing mechanism 
 is generally expected to occur due to strong electron correlation between heavy quasiparticles.
 The discovery of heavy-fermion superconductivity in UBe$_{13}$~\cite{Ott_PRL_1983}  triggered exploration of  
 unconventional pairing mechanism  in $5f$ actinide compounds, and subsequently two uranium compounds, UPt$_3$~\cite{Stewart_PRL_1984} and URu$_2$Si$_2$~\cite{URS1,URS2}, were found to show  superconductivity. 
 These U-based heavy-fermion superconductors have attracted considerable interest because of their unusual 
 superconducting (SC) and normal-state properties.  
Among these, superconductivity in UBe$_{13}$ is \correct{highly enigmatic; it emerges from a strongly non-Fermi-liquid state with a} large resistivity ($\rho \sim 150$ $\mu \Omega$cm).
Also unusual is the temperature variation of the upper critical field $H_{\mathrm{c2}}$: an enormous initial slope $-(dH_{\rm c2}/dT)_{T_c}\sim 42$~T/K and an apparent absence of a Pauli paramagnetic limiting at low temperatures~\cite{Maple85}.
\correct{Extensive studies have been made to elucidate the SC gap symmetry~\cite{MacLaughlin_PRL_1984,Walti_PRBR_2001}, with an expectation of an odd-parity pairing in this compound~\cite{Ott_PRL_1984,Einzel86,Tien_PRB_1989,Fomin00}.}
Recently, it has been found quite unexpectedly that nodal   quasiparticle excitations in UBe$_{13}$ are absent
  as  revealed by  low-$T$ angle-resolved heat-capacity measurements for  a single crystalline sample
 \cite{YShimizu_PRL_2015}. 

A long-standing mystery regarding  UBe$_{13}$   is  the occurrence of  \correct{a second} phase transitions in the SC state when a small amount of Th  is \correct{substituted}  for U \correct{ [Fig.~1(a)]}
 \cite{Smith_Physica_1985, Ott_PRBR_1985}. 
 It has been reported that there exist four phases (A, B, C, and D) in its SC state, according to the previous $\mu$SR 
\cite{Heffner_PRL_1990} and thermal-expansion 
  \cite{Kromer_PRL_1998} experiments using polycrystalline samples.
\correct{The  SC transition temperature $T_{\rm c}$ is non-monotonic as a function of the Th concentration $x$ in U$_{1-x}$Th$_{x}$Be$_{13}$, and exhibits a sharp minimum near $x=0.02$. Further doping of Th results in an increase of the bulk SC transition temperature ($T_{\rm c1}$), reaching a local maximum at $x\sim 0.03$}~\cite{Smith_Physica_1985}.
Below $T_{\rm c1}$,  another \correct{phase transition accompanied by a} large heat-capacity jump occurs  at $T_{ \rm c2}$ \correct{in a narrow} range of 0.019$<$ $x$ $<$0.045~\cite{Ott_PRBR_1985, Heffner_PRL_1990}.
Interestingly,  only for this $x$ region, 
weak \correct{magnetic correlations} have been observed in zero-field $\mu$SR measurements
 \cite{Heffner_PRL_1990}.
\correct{The} previous  thermal-expansion study 
 \cite{Kromer_PRL_1998}
\correct{claimed} that 
\correct{the low-temperature (``$T_{\rm L}$'') anomaly appearing below $T_{\rm c}$ for $0\leq x<0.02$, which can be connected to} the ``$B^*$ anomaly'' observed  in pure UBe$_{13}$ \cite{Kromer_PRL_1998, Ellman_PRBR_1991, YShimizu_PRL_2012},
 is a  precursor of the transition at $T_{\rm c2}$.
Up to present, the true nature of the transition at $T_{\rm c2}$ remains controversial
 \cite{Steglich_RPP_2016, Kenzelmann_RPP_2017}.
Two \correct{different} scenarios  have been discussed so far on the $T_{\rm c2}$ transition: 
  (i)  an additional  SC transition  that \correct{breaks} time-reversal symmetry
  \cite{Sigrist_RevModPhys_1991},
 and  
 (ii) \correct{the occurrence of an antiferromagnetic ordering that coexists with the} SC state~\cite{Batlogg_PRL_1985, Machida_PRL_1987}.
 Indeed,   although 
 it has been reported that  the NMR spin-relaxation rate 
 \cite{MacLaughlin_PRL_1984},  heat capacity 
 \cite{Jin_PRBR_1994},
 and muon Knight shift 
  \cite{Sonier_muSR}, 
 show unusual  temperature dependence in the SC state,  
  little is known concerning  the gap structure in U$_{1-x}$Th$_{x}$Be$_{13}$ due to the lack of information about the \textit{ anisotropy }   of  its quasiparticle excitations in magnetic fields. 
\color{black}

\begin{centering}
\begin{figure}[t]
\begin{minipage}{4.0cm}
\includegraphics[height=4.6cm]{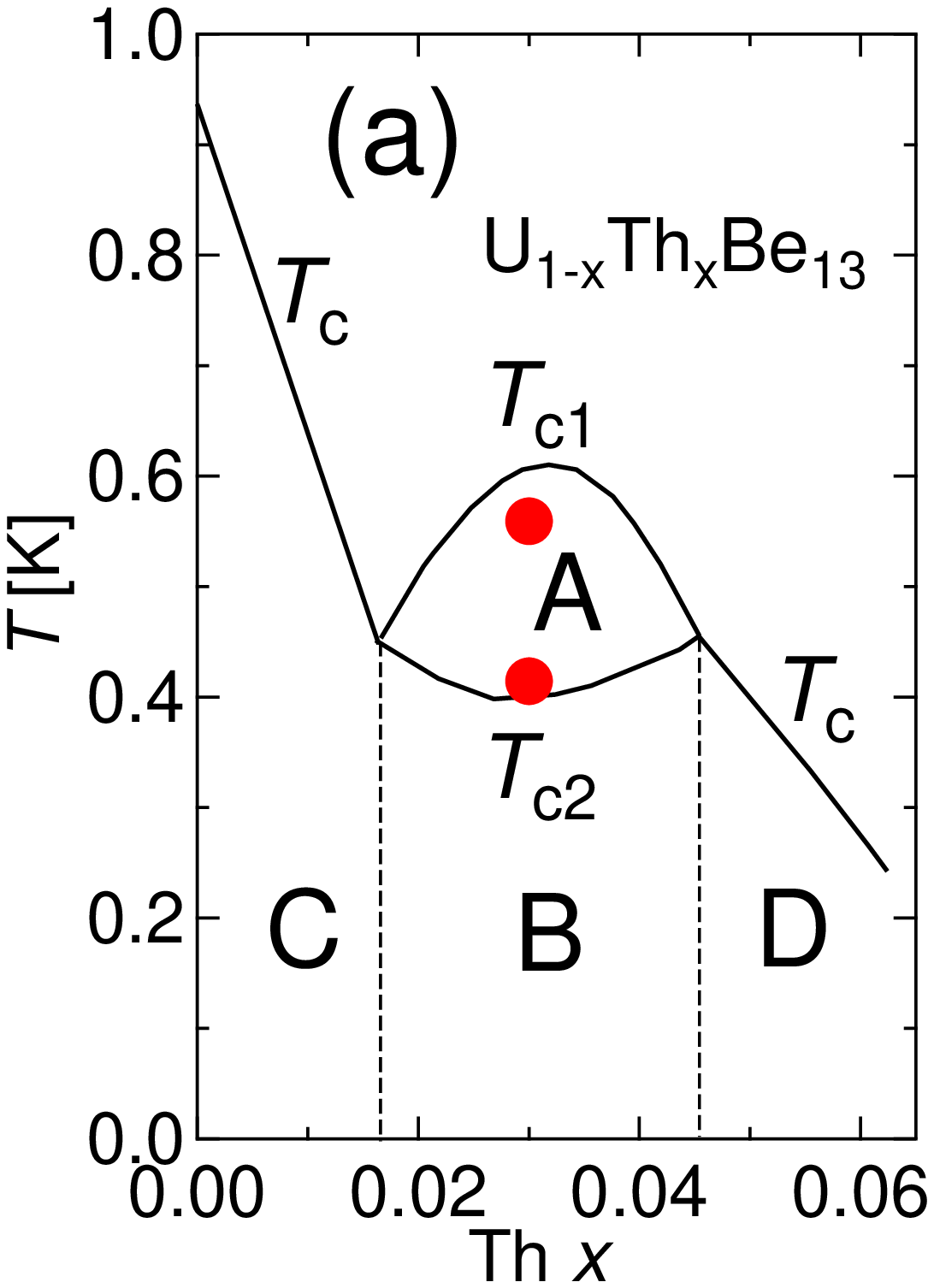}
\end{minipage}
\begin{minipage}{4.0cm}
\includegraphics[height=2.3cm]{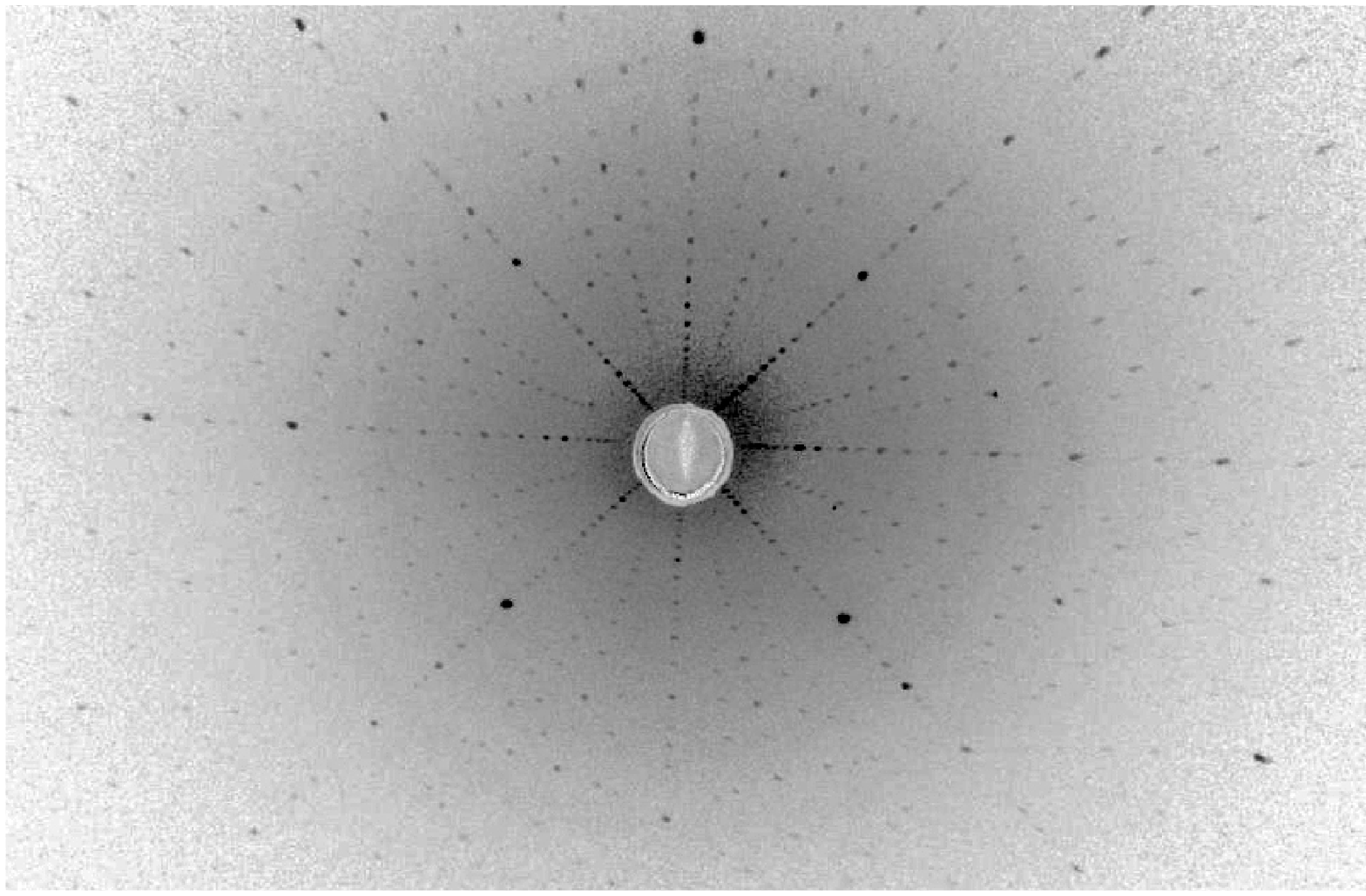}
\includegraphics[height=2.3cm]{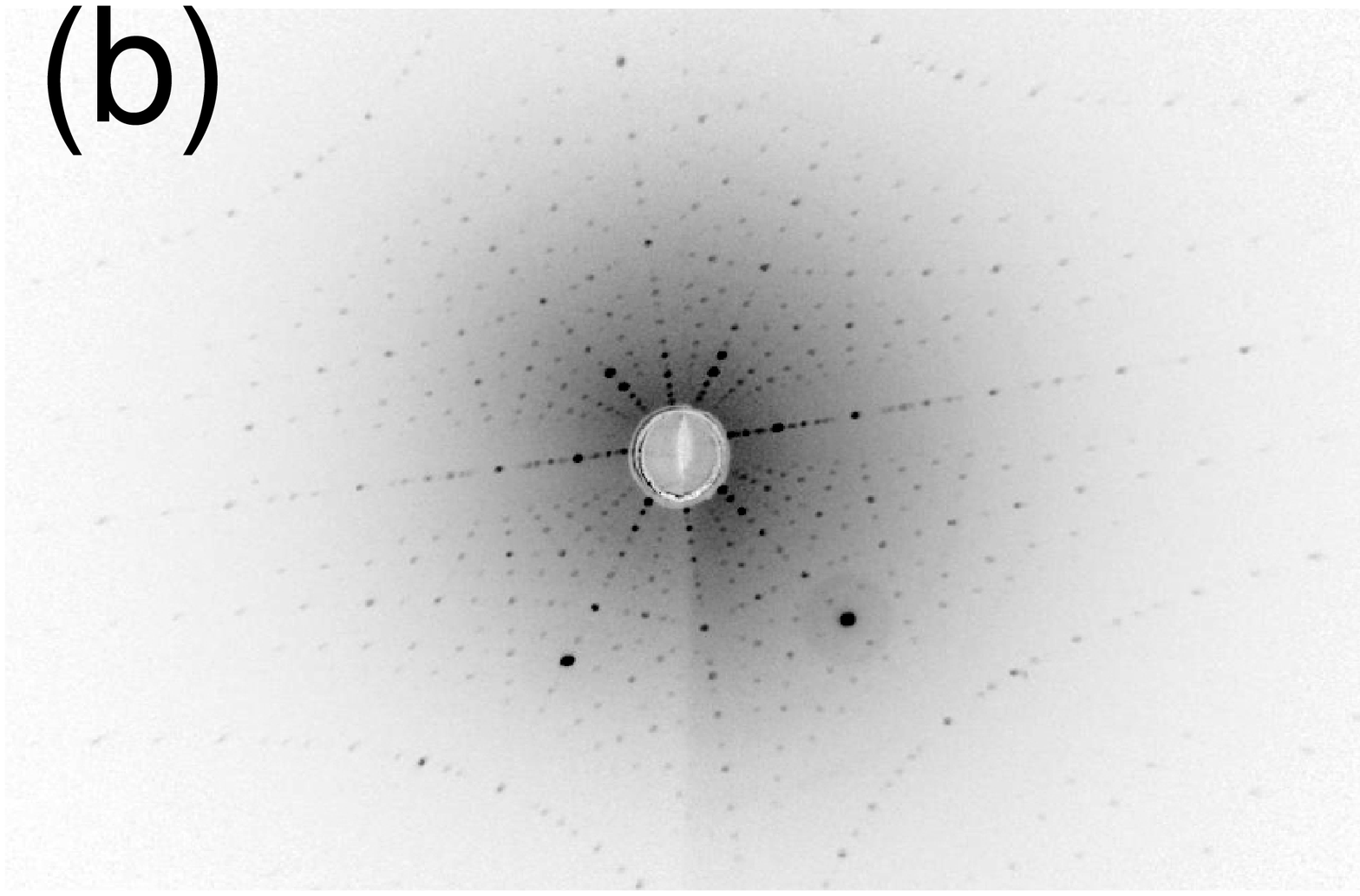}
\end{minipage} 
\includegraphics[width=8.6cm]{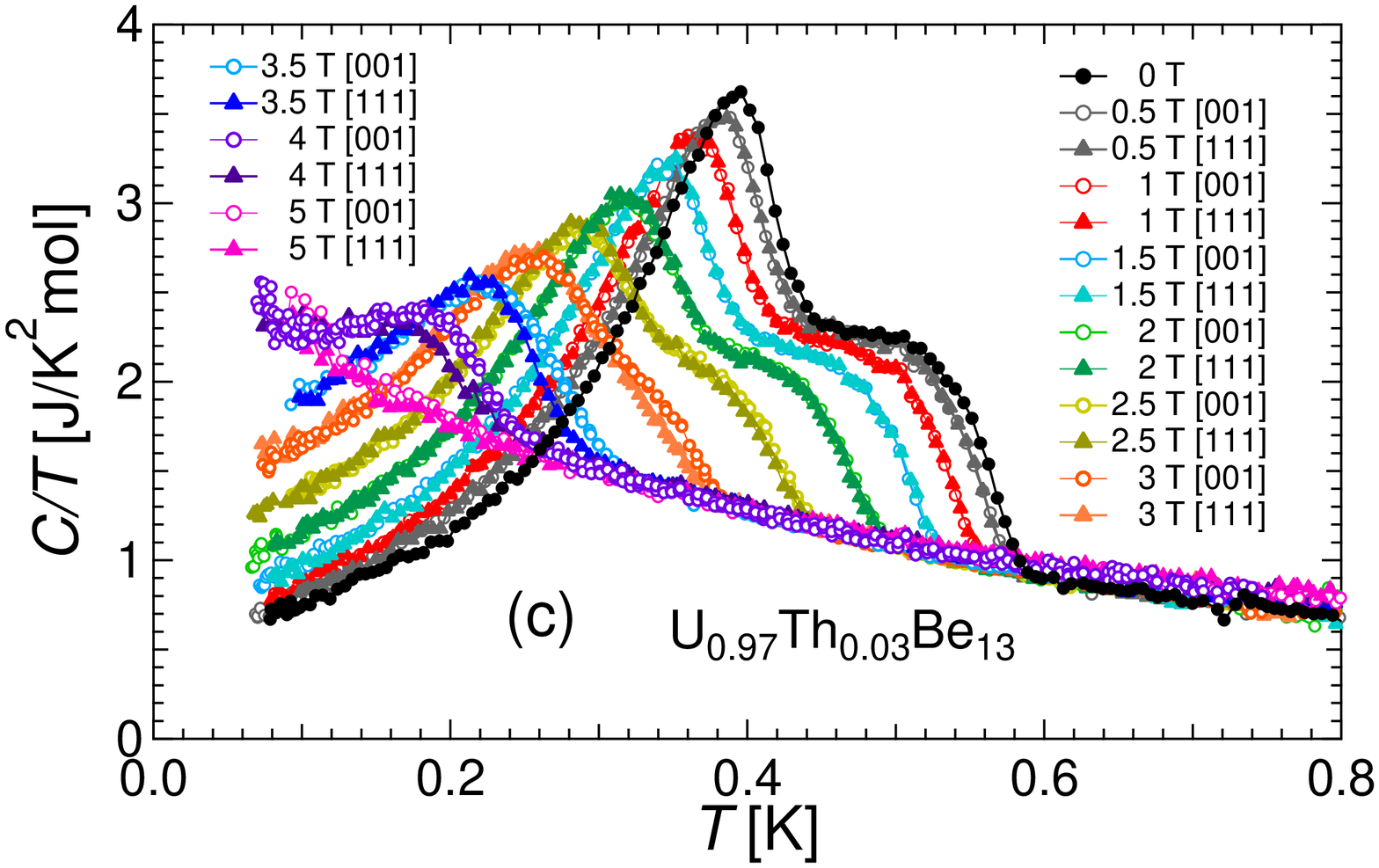}
\caption{   (Color online)
 (a) 
 $T$-$x$ phase diagram of U$_{1-x}$Th$_{x}$Be$_{13}$ \cite{Heffner_PRL_1990, Kromer_PRL_1998}, where the solid lines are based on 
 ref. \cite{Kromer_PRL_1998}.
   There are  four phases (A, B, C, and D) in its SC state, according to the previous 
 $\mu$SR \cite{Heffner_PRL_1990} and thermal-expansion \cite{Kromer_PRL_1998} studies.
 The red circles
 indicate the transition temperatures  at zero field for the sample used in the present  experiment ($x$ = 0.03). 
Here  $T_{\rm c1}$ and $T_{c2}$ are 
 determined from $C(T)$ by considering the entropy conservation at each transition.
\color{black}
  (b) Laue X-ray photographs for  the cubic 4-fold $(100)$ \correct{(upper panel)} and    2-fold $(1\bar{1}0)$  \correct{(lower panel)} planes. 
 (c)  $C(T)/T$ \correct{of U$_{0.97}$Th$_{0.03}$Be$_{13}$} 
 at zero and \correct{in} magnetic fields up to 5 T, \correct{measured every 0.5 T step for two directions $H\parallel [001$] (circles) 
 and $H\parallel [111]$ (triangles)}.
 }
\end{figure}
\end{centering}

\color{black}


In order to  resolve the controversy 
  regarding the second transition  
 at $T_{\rm c2}$,  
and to uncover its gap symmetry, 
  in this Letter we report the results of high-\correct{precision} heat-capacity and dc magnetization measurements \correct{on} 
  U$_{0.97}$Th$_{0.03}$Be$_{13}$.
Single-crystalline U$_{0.97}$Th$_{0.03}$Be$_{13}$   samples were obtained using a  tetra-arc furnace;  
  the ingot was remelted several times and then quenched.
\correct{By} this procedure, we have succeeded in obtaining  small monocrystalline samples  with no additional heat treatment  
 as confirmed by sharp X-ray Laue spots in Fig.~1(b).
\color{black}
 Heat capacity ($C$)    was measured  at low temperatures  down to $60$ mK  by means of a standard quasi-adiabatic heat-pulse method  in a $^3$He-$^4$He dilution refrigerator.
Field-orientation dependences  $C(H, \phi)$
 were \correct{obtained}  under rotating magnetic fields in the $(1\bar{1}0)$ crystal plane  that includes the  $[001]$, $[111]$, and $[110]$ \correct{axes},  using \correct{a} 5T$\times$3T vector magnet.    \correct{We define} the angle $\phi$  measured from the $[001]$ \correct{direction}.
Dc magnetization measurements were performed \correct{along the $[1\bar{1}0]$ axis} down to 
 $T\sim$0.28 K  for the same single crystal  using  a capacitive Faraday magnetometer 
 \cite{Sakakibara_JJAP_1994}  installed \correct{in} a $^{3}$He refrigerator.
A magnetic-field gradient of 9 T/m was  applied to the sample, independently of the central field at the sample position.


Figure 1(c) shows $C(T)/T$ curves measured   at zero and various   fields up to 5 T   applied along   $[001]$ and $[111]$ axes.
At \correct{zero field},  two \correct{prominent}  jumps   
 occur  at $T_{\rm c1} \sim$0.56 K and $T_{\rm c2} \sim$0.41 K, 
 \correct{where the transition temperatures 
 [red circles in Fig. 1(a)]  
 are determined by transforming the broadened transitions into idealized sharp ones by an equal-areas construction.} 
\correct{The results are in agreement with the previous reports}~\cite{Smith_Physica_1985,Ott_PRBR_1985}.
With increasing field, \correct{both transitions shift to lower temperature,  getting closer to each other~\cite{Ott_PRBR_1986,Kromer_PRB_2000}.}
Above 3.5 T, 
\correct{the} two transitions \correct{become so close to each other and are difficult to resolve separately}.
\correct{There is a notable feature in the}   anisotropy of $C(T)$ in magnetic fields. 
\correct{At low fields} below $\sim$1.75 T,  \correct{the shifts of the two}  transition temperatures  are 
 \correct{almost} isotropic.
\correct{At higher fields above 2.5~T},  however,  the $T_{\rm c1}(H)$  becomes \correct{slightly} anisotropic, \correct{$T_{\rm c1}(H||[001])>T_{\rm c1}(H || [111])$},
\correct{while}  $T_{\rm c2}(H)$  remains isotropic.
 In general, an anisotropy of $T_{\rm c}(H)$ and $H_{\rm c2}$  
 results from those of SC gap function and/or   Fermi velocity.
 If the  double trasnitions come from two inhomogeneous SC states with the same gap symmetry,
 they should show the same anisotropic (or isotropic) field response. 
 Our experimental results exclude  such an extrinsic possibility. 
 Thus
  the difference between field anisotropy in 
  $T_{\mathrm{c1} }(H)$ and $T_{\mathrm{c2} }(H)$ is an essential 
 effect which strongly suggests that the order parameters of these two phases have qualitatively different field-orientation dependences.
\color{black}

\begin{centering}
\begin{figure}
\includegraphics[width=8.5cm]{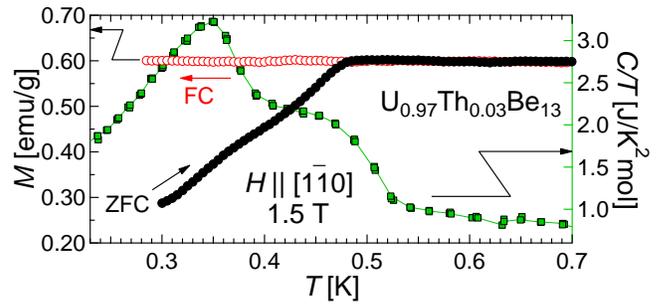}
\caption{ (Color online)  \correct{Temperature} dependence of \correct{the} dc magnetization  $M(T)$ measured at 1.5 T
  for $H$$||$$[1\bar{1}0]$.  The data of $C(T)/T$ \correct{measured in the same magnetic field are also plotted for comparison}.
   }
\end{figure}
\end{centering}

A  key question, then,  is  whether the second transition at $T_{\rm c2}$ is a SC transition into a different gap symmetry.
 To address this question, we performed precise dc magnetization [$M(T)$] measurements  across the double transitions.
Figure 2 shows the temperature dependence of  $M(T)$ measured at 1.5~T \correct{together} with the $C(T)/T$ data for the same \correct{field on the same} sample.
\correct{FC and ZFC denote the data taken in the field-cooling and zero-field-cooling protocols, respectively.}
\correct{The FC-ZFC branching occurs below $\sim 0.5$~K close to $T_{\rm c1}$ at this field, indicating the appearance of bulk superconductivity.}
\correct{We find a small but distinct kink in the ZFC data near $T_{\rm c2}$, while no such anomaly can be seen in the FC curve. This fact implies a substantial change in the vortex pinning strength at this temperature,  consistent with the previous vortex creep measurements~\cite{Zieve95,Mota2000}. }
\correct{Regarding the possible origin of the enhanced flux pinning in the low-$T$ phase, we find no signatures that can be ascribed to a magnetic transition in the FC curve near $T_{\rm c2}$.}
\correct{Our magnetization data, therefore, strongly suggest that the transition at $T_{\rm c2}$ is of a kind such that the SC order parameter changes. Indeed, it has been argued that such an enhancement of the vortex pinning occurs in a SC state with broken time reversal symmetry~\cite{Zieve95}.}
This conclusion is also consistent with  \correct{the previous} neutron scattering measurements~\cite{Hiess_PRB_2002} 
 which show no evidence for magnetic ordering \correct{in U$_{0.965}$Th$_{0.035}$Be$_{13}$ down to 0.15~K}.

\color{black}

Next we  examine the magnetic-field dependence of \correct{the} heat capacity and its anisotropy \correct{in more detail}, whose  behavior  in low fields reflects quasiparticle excitations  \correct{in} the SC \correct{state and provides a hint for the gap symmetry}~\cite{Volovik_JETP_Lett_1993, Vekhter_PRBR_1999, Sakakibara_ROPP_2016}.
Figure 3(a)
 shows  $C(H)/T$ measured at \correct{$T=$} 0.12, 0.18, 0.24, 0.30, 0.36, and 0.40 K  for  the cubic $[001]$ and $[111]$ directions, and the inset shows the enlarged $C(H)/T$ \correct{plot obtained}  at  0.08 K.
Note  that $C(H)$  \correct{below 1~T} is \correct{quite} linear to $H$ at \correct{the} lowest temperature \correct{of} 0.08 K.
This  \correct{behavior is in striking} contrast with a convex upward $H$ dependence  \correct{expected} for nodal superconductors~\cite{Nodal_HeatCapacity}.
Moreover,  there is \textit{no} anisotropy in  $C(H) \propto H$ between $H$ $||$  $[001]$ and $[111]$ in low fields below $\sim$2 T.
The absence of the anisotropy is further demonstrated by \correct{angle-resolved} $C(\phi)/T$
 \correct{in Fig. 4(a),}
  obtained in a field of 1~T rotated in the $(1\bar{1}0)$ crystal plane at $T=$0.08, and 0.42~K, \correct{together with the result} measured  in the normal state at 0.60 K.
\correct{The absence of any angular dependence in $C(\phi)/T$ in a low-$T$ low-$H$ region again excludes  the possibility of a nodal-gap structure in which a characteristic angular oscillation should be expected in $C(\phi)/T$~\cite{Vekhter_PRBR_1999}.}
The \correct{present} $C(H, \phi)$ data \correct{thus indicate that  nodal quasiparticles are absent}
  in  U$_{0.97}$Th$_{0.03}$Be$_{13}$, similarly to the behaviors \correct{observed} in  pure UBe$_{13}$ 
 \cite{YShimizu_PRL_2015}.

\begin{centering}
\begin{figure}
\includegraphics[width=8.7cm]{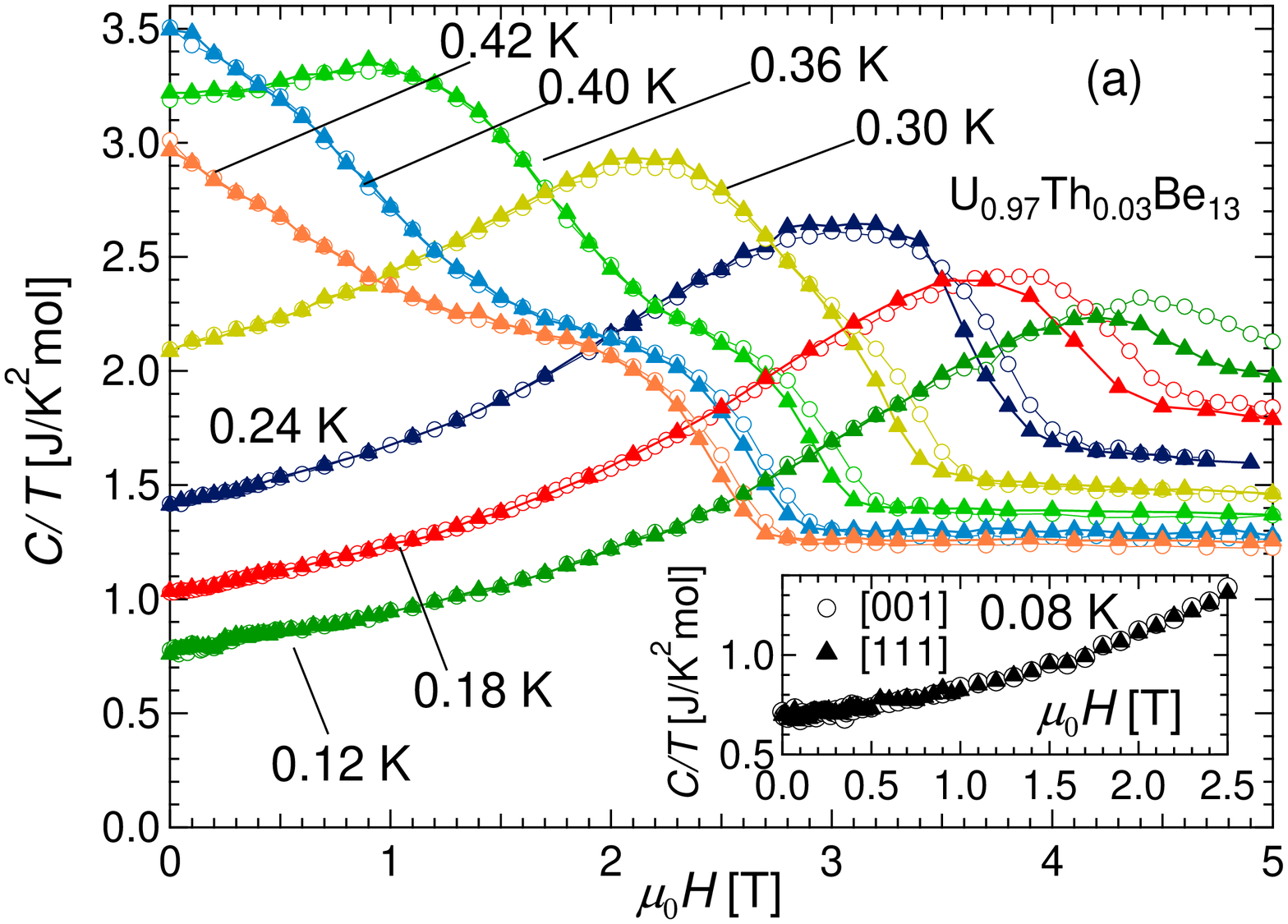}
\begin{minipage}{4.1cm}
\includegraphics[width=4.0cm]{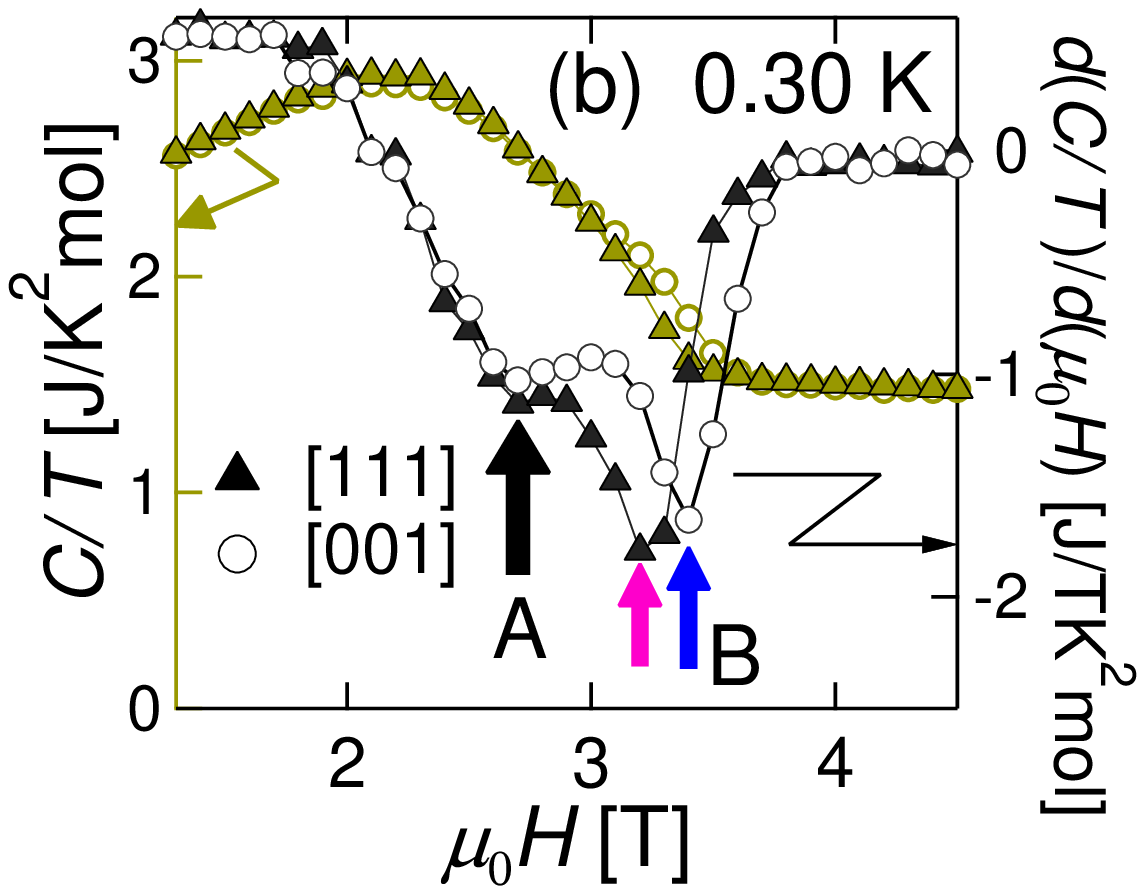}
\end{minipage}
\begin{minipage}{4.1cm}
\includegraphics[width=4.0cm]{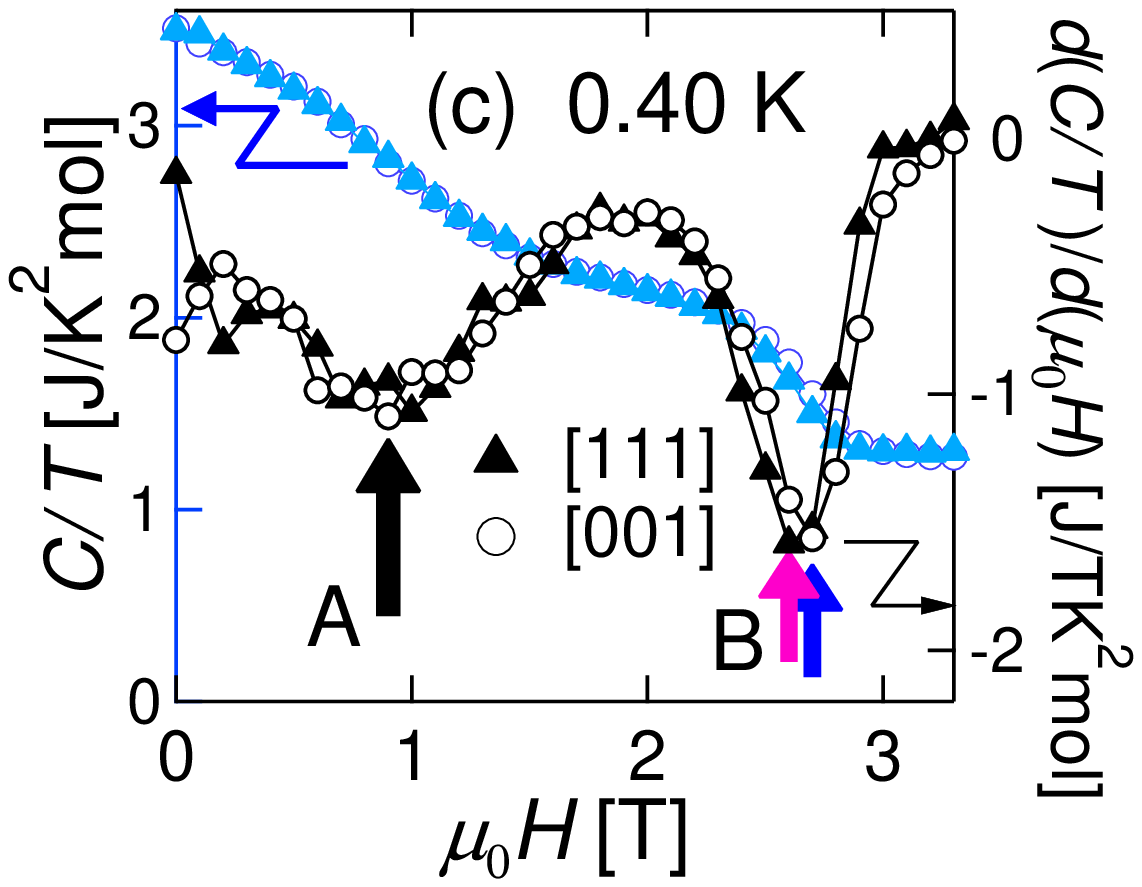}
\end{minipage} 
\caption{ (Color online) (a) Magnetic-field dependence of  $C(H)/T$ up to 5~T
for $H$$||$$[001]$ (circles) and $H$$||$$[111]$ (triangles) measured at $T=$ 0.12, 0.18, 0.24, 0.30, 0.36, 
 0.40, and 0.42~K. 
 The inset shows the  $C(H)/T$ in low magnetic fields measured at  the \correct{base} temperature of  $T= $0.08 K. 
 $C(H)/T$ and its differential as a function of magnetic field around the double transitions at 
 (b) 0.30 and (c) 0.40 K. The transition fields of the A and B phases, \textit{i.e.}, 
 $H_{\rm c2}^{\rm A}$  and  $H_{\rm c2}^{\rm B}$,  are determined as magnetic fields where the differential, 
 $ d[C(H)/T]/dH$, shows a local minimum.
}
\end{figure}
\end{centering}

At higher fields, double-step-like anomalies are observed  in $C(H)/T$ at 0.42, 0.40, and 0.36 K 
 [Fig. 3(a)].
Here the double transitions can be  cleary defined by the differential data, $d[C(H)/T]/dH$, 
  as shown in Figs. 3(b) and 3(c).
The lower-field \correct{step} occurs when \correct{the boundary} $T_{\rm c2}(H)$ \correct{is crossed},
 while the higher-field one  corresponds to the transition at  $T_{\rm c1}(H)$, \textit{i.e.}, the upper critical field 
 $H_{\rm c2}(T) \equiv H_{\rm c2}^{\rm A}$.
Note that the position of the lower-field anomaly 
 ($H_{\rm c2}^{\rm B}$)
 \color{black}
  is fully \textit{isotropic}, 
 whereas  the higher-field one 
 ($H_{\rm c2}^{\rm A}$)
 \color{black}  
shows an appreciable anisotropy, indicating that $H_{\rm c2}$ becomes anisotropic: 
 $H_{\rm c2}^{\rm A}  \parallel [001] > H_{\rm c2}^{\rm A} \parallel [111]$.
The anisotropy of 
 $H_{\rm c2}^{\rm A}$ 
becomes larger  at lower temperatures.
With decreasing $T$,  \correct{both of the transition fields shift to higher fields, getting close to each other,} and \correct{are difficult to discriminate} below  $\sim$0.24 K 
 [Fig. 3(a)].
   \color{black} 
\correct{These features of the transition fields} are fully consistent with \correct{those observed} for $T_{\rm c1}(H)$ and $T_{\rm c2}(H)$
\correct{shown in Fig.~1(c)}.
Note that   the  isotropic behaviors  in   $C(H)/T$ as well as $T_{\rm c2}(H)$ (Fig. 3)  
  contrast starkly with the anisotropic  behavior of  $B^*$ anomaly found  in pure UBe$_{13}$ 
 \cite{YShimizu_PRL_2015}, 
 suggesting that these phenomena may result from different origins. 
Figure 4(b) shows  the $H$-$T$ phase diagram of  U$_{0.97}$Th$_{0.03}$Be$_{13}$ \correct{determined from the present $C(T,H)$ measurements, where the two SC phases are denoted as A and B phases. 
The overall features of the phase diagram 
 are \color{black}
 essentially the same with those obtained previously~\cite{Ott_PRBR_1986,Kromer_PRB_2000,Jin_PRB_1996}}.
In Fig.~4(a), $C(\phi)/T$ data measured at  $T=0.36$~K in  $\mu_{0}H=3$~T  (A phase) rotated in the $(1\bar{1}0)$ are also shown; 
  $C(\phi)/T$ shows a distinct angular oscillation with the maximum (minimum) along the [001] ([111]) direction, reflecting the anisotropy in
     $H_{\rm c2}^{\rm A}$.
  \color{black}

\begin{centering}
\begin{figure}
\begin{minipage}{4.1cm}
\includegraphics[height=5.2cm]{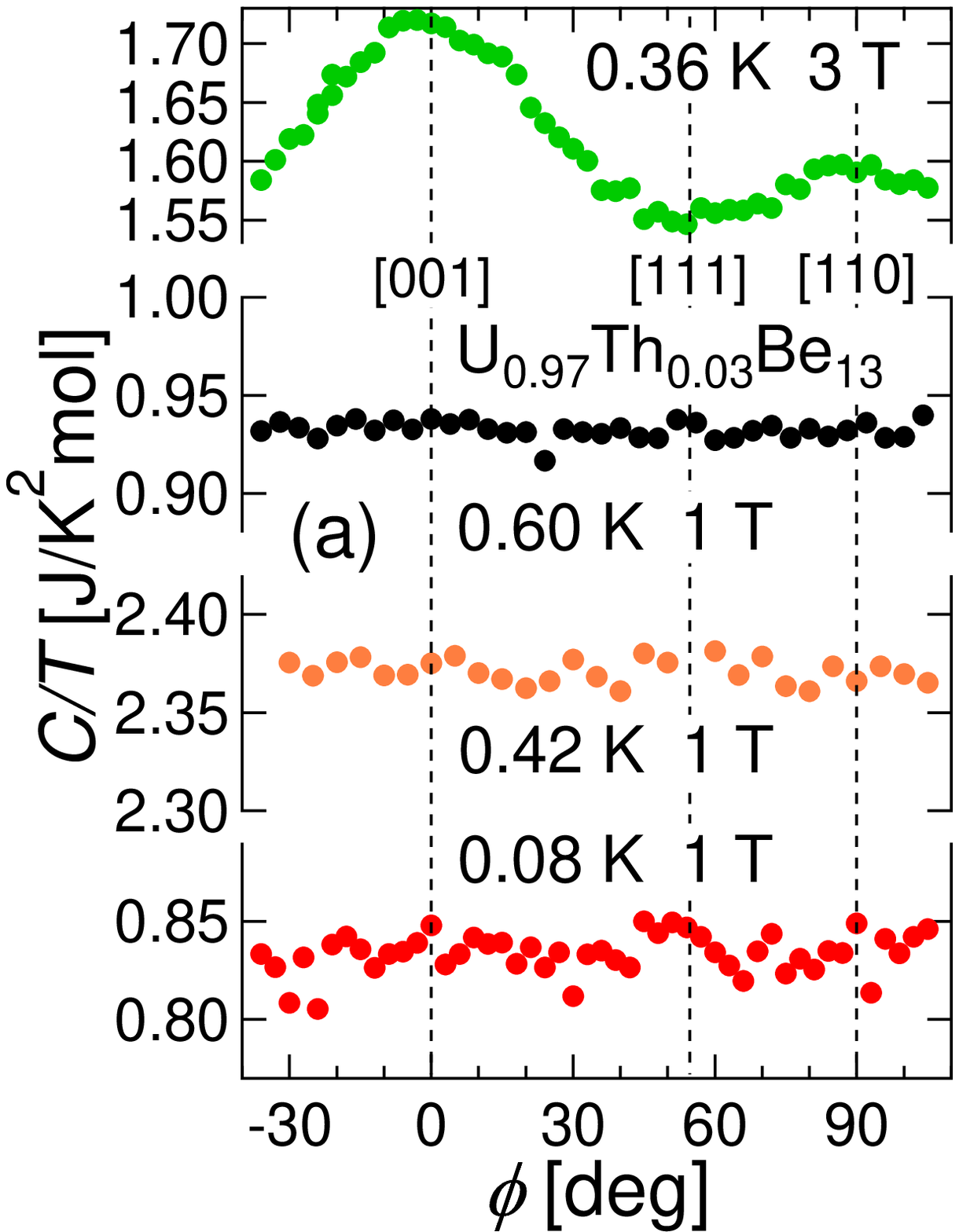}
\end{minipage}
\begin{minipage}{4.1cm}
\includegraphics[height=5.2cm]{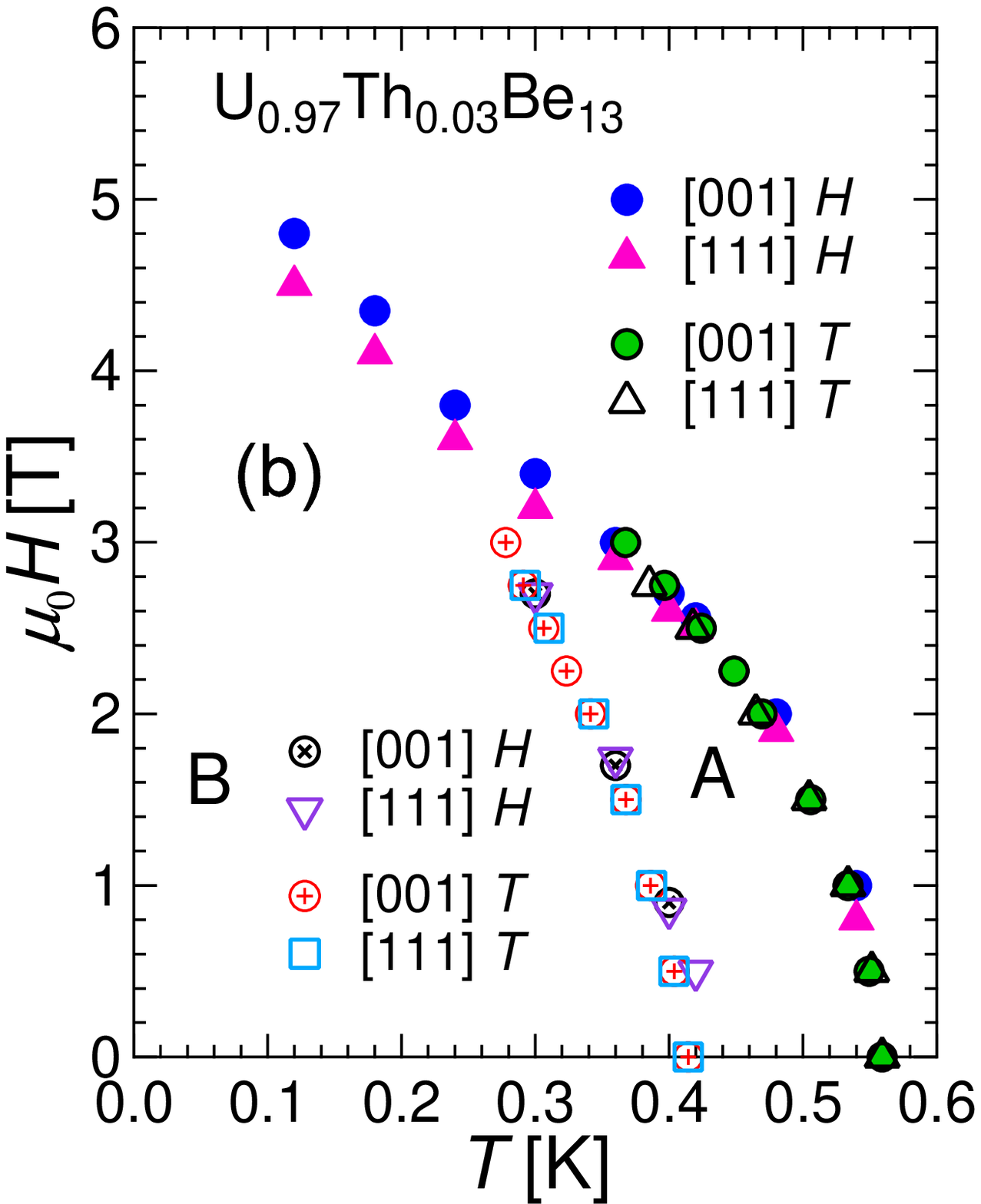}
\end{minipage} 
\caption{ (Color online) 
(a) Angular dependence of $C(\phi)/T$, measured at $T=$0.08 (B phase), 0.42 (A phase), and 0.60 K (normal state),   
in a magnetic field of  1~T.
  $C(\phi)/T$, measured at $T=$0.36 K in 3 T  (A phase), near $H_{\mathrm{c2} }$ is also plotted.   
(b)  
 $H$-$T$ phase diagram for the SC state of U$_{0.97}$Th$_{0.03}$Be$_{13}$ for $[001]$ and $[111]$, 
 where  $T$ and $H$ denote data obtained from temperature and field scans, respectively.
Here, $T_{\mathrm{c1} }$ and $T_{\mathrm{c2} }$ were determined by considering entropy conservation
 at  transitions  in the $C(T)/T$ curves.
 }
\end{figure}
\end{centering}

\color{black}

The present experiment thus provides strong evidence that U$_{0.97}$Th$_{0.03}$Be$_{13}$ exhibits
  double SC transitions \color{black} with two different SC order parameters.
Let us  discuss   possible SC gap symmetries in \correct{this system}.
 A key experimental fact is that the SC gap is fully open over the Fermi surface in both the B and C phases,
 as suggested by the present and previous \cite{YShimizu_PRL_2015} studies, respectively.
\color{black}
\correct{This would imply either}
 (i) the SC gap function itself \correct{to be} nodeless, or 
 (ii) the SC gap function \correct{to have} nodes only in the directions in which the Fermi surface is missing. 
\correct{Regarding the latter,} band calculations \correct{tell us that}
 the Fermi surface is missing along the $\langle 111\rangle$ direction, except for a tiny electron band~\cite{Takegahara_PhysicaB_2000, Maehira_PhysicaB_2002}.
Given the fact that spontaneous magnetism is observed from zero-field $\mu$SR only
  below $T_{\rm c2}$~\cite{Heffner_PRL_1990}, \correct{in addition},
 it \correct{would be natural to assume} that the B phase is a time-reversal-symmetry broken SC state.
\correct{Under these constraints,} two plausible scenarios \correct{can be proposed to explain the multiple SC phases in U$_{1-x}$Th$_{x}$Be$_{13}$.}
One  is \correct{to employ a degenerate order parameter} 
  belonging to \correct{higher dimensional} representations of  the $O_{h}$ symmetry  (degenerate scenario). 
The other is  to assume  
two \correct{order parameters belonging to} different representations of the $O_{h}$ group,   \correct{nearly degenerate to each other} (accidental scenario)
 \cite{Sigrist_RevModPhys_1991}.

\textit{Degenerate scenario: } 
The group theoretic classification \correct{of the gap functions} under the  cubic symmetry $O_{h}$ has been given by several authors~ \cite{Volovik_JETP_1985, Blount_PRB_1985, Sigrist_RevModPhys_1991,Ozaki_Machida_Ohmi_PTP_1985}.
Among them, the two-dimensional  odd-parity  $E_{u}$ state is  a promising candidate \correct{for the order parameter} which naturally explains \correct{the existing} experimental data of both pure and Th-doped  UBe$_{13}$
 \cite{Endnote_Eg_symmetry}. 
The possibility of  the odd-parity state has  also been suggested from the $\mu$SR  Knight shift  experiments
  \cite{Sonier_muSR}.
\color{black}
\correct{As for}  the odd-parity $E_{u}$ state,
\correct{we have} two basis  functions, $\bm{l}_{1} (k) = \sqrt{3} (\hat{ \bm{x} } k_{x} - \hat{ \bm{y} } k_{y}  )$, and
 $\bm{l}_{2} (k) = 2 \hat{ \bm{z} } k_{z} - \hat{ \bm{x} } k_{x}  - \hat{ \bm{y} } k_{y}   $, 
 and their  combinated state, 
 $\bm{d (k)} = \bm{l}_{1} + i \bm{l}_{2} =  \hat{ \bm{x} } k_{x} + \epsilon \hat{ \bm{y} } k_{y} +  \epsilon^2 \hat{ \bm{z} } k_{z} $
 with $\epsilon = e^{i \frac{2 \pi }{3} }  (\epsilon^3 = 1)$. 
The non-unitary state  $\bm{d (k)} = \bm{l}_{1} + i \bm{l}_{2} $ has  point  nodes only   along  the $\langle111\rangle$ direction, 
  therefore,  the nodal  quasiparticle excitations can be missing
  considering the \correct{calculated Fermi surface} 
 \cite{Takegahara_PhysicaB_2000, Maehira_PhysicaB_2002}.
The condition of the occurrence of each  two-dimensional SC state can be examined using the Ginzburg-Landau free energy density,
 $F = \alpha(T) (| \bm{l}_{1} |^2 + | \bm{l}_{2} |^2 ) + \beta_{1} (| \bm{l}_{1} |^2 + | \bm{l}_{2} |^2 )^2  + \beta_{2} ( \bm{l}_{1} \bm{l}^{*}_{2} +  \bm{l}^{*}_{1} \bm{l}_{2}  )^2 $ with 
  $\alpha(T) = \alpha_{0} (T_{c} -T ) $,
 where $\beta_{1} > 0$ is required for the stability. 
If $\beta_{2} > 0$, the non-unitary  state with the broken time-reversal symmetry becomes stable in lower $T$  
 as a ground state (the B phase).
With increasing temperature the degeneracy of the order parameters is lifted at $T_{\rm c2}$,
  and one of them appears in 
 the A phase ($T_{\rm c2}$ $<$ $T$ $<$ $T_{\rm c1}$).
Logically, 
 the other one appears in the C phase by changing dopant $x$.
In  pure UBe$_{13}$ (the C phase),   a nodeless gap function, \textit{i.e.}, 
 $\bm{l}_{2} (k) = 2 \hat{ \bm{z} } k_{z} - \hat{ \bm{x} } k_{x}  - \hat{ \bm{y} } k_{y} $, which is a unitary state, 
  is likely, 
  explaining the absence of nodal quasiparticle excitations~\cite{YShimizu_PRL_2015}
\correct{without invoking} the Fermi-surface topology.

\textit{Accidental  scenario: } 
We  briefly discuss the possibility of the accidental scenario, starting with   
 the simplest and most symmetric  $A_{1u}$,  namely $\bm{d}_{A1u} (\bm{k} ) =\hat{\bm {x} } k_{x} +\hat{\bm {y} } k_{y} +\hat{\bm {z} } k_{z} $ with an isotropic full gap  as the C phase for $x$ = 0.
From $x$ = 0.019 to $x$ = 0.045, we consider the combined state of 
 1D representations, the above  $p$-wave  $A_{1u}$ and $f$-wave $A_{2u}$ with  $\bm{d}_{A2u} (\bm{k} ) =\hat{\bm {x} } k_{x} (k_{y}^2 - k_{z}^2)  + \hat{\bm {y} } k_{y} (k_{z}^2 - k_{x}^2 ) + \hat{\bm {z} } k_{z} (k_{x}^2 - k_{y}^2) $.
The combined state  of  $A_{1u}$ and $A_{2u}$, namely,  non-unitary $ \bm{d} (\bm{k}) = \bm{d}_{A1u} + i \bm{d}_{A2u}$ is \correct{nodeless irrespective}  of the Fermi-surface topology, although  $\bm{d}_{A2u} $  alone  has point nodes  along  $\langle100\rangle$ and   $\langle111\rangle$ directions.
Thus nodeless $A_{1u}$ and the  $A_{1u} + iA_{2u}$ states  can explain the 
  absence of nodal quasiparticles in  pure and Th-doped UBe$_{13}$, respectively
 \cite{Endnote_isotropic_A1g}.
Similarly, the other order parameters belonging to different irreducible representations are possible, \textit{e.g.}, $A_{1u} + i E_{u}$;
  the determination of the two order parameters is not easy due to the arbitrariness of their combinations.

Finally,  it is worth  discussing  the topology of the $H$-$T$ phase diagram.
In Fig.~4(b),  \correct{it may appear} that  the lines of $T_{\rm c1}(H)$ and $T_{\rm c2}(H)$
 merge \correct{into a single 2nd-order transition line} in a high-field region.
\correct{Such a case is,}  however, \correct{not allowed in the thermodynamic argument of the multicritical point~\cite{Yip91PRB,Endnote_1stOrderPhaseTransition}}.
\correct{Instead,}  a crossing of the two  \correct{2nd-order} transition lines at a \correct{tetra}-critical point is possible
 \cite{Yip91PRB}.
This argument \correct{imposes the existence of} another  2nd-order transition  below $H_{\rm c2}$
 for $T \lesssim 0.25$ K, but \correct{no evidence for such a transition line has  been obtained so far in our measurements as well as in previous thermal expansion studies~\cite{Kromer_PRL_1998}}. 
\correct{It might be natural to consider an anti-crossing of the two 2nd-order transition lines~\cite{Machida_JPSJ_1989}.}
The crossing of $T_{\rm c1}(H)$ and $T_{\rm c2}(H)$ in U$_{1-x}$Th$_{x}$Be$_{13}$  will be examined further in future studies. 

\color{black}

To conclude, 
  low-energy quasiparticle excitations and magnetic response  of U$_{0.97}$Th$_{0.03}$Be$_{13}$  were studied by means of   heat-capacity and dc magnetization measurements.
The  magnetization results  evidence  that the \correct{second} transition at   $T_{\mathrm{c2}}$
 is between two different 
 SC 
 states.
Strikingly, the present $C(T,H,\phi)$ data
 strongly suggest
   that the SC gap is fully open over the Fermi surface in U$_{0.97}$Th$_{0.03}$Be$_{13}$,
 excluding a number of   gap functions possible in the cubic symmetry.
Our new thermodynamic results  
  entirely  overturn a widely believed idea that 
  nodal quasiparticle excitations occur in the odd-parity SC state with  broken time-reversal-symmetry.
The absence (presence) of anisotropy for $T_{\rm c2}$ ($T_{\rm c1}$)
    in  fields clearly demonstrates  that the gap symmetry in the B phase ($T$ $<$ $T_{\rm  c2}$)   is 
   distinguished  from that of the A phase ($T_{\rm c2}$ $<$ $T$ $<$ $T_{\rm c1}$).
 Moreover, the isotropic behavior of the $T_{\rm  c2}(H)$ in U$_{1-x}$Th$_{x}$Be$_{13}$ 
  contrasts starkly to  the anisotropic field response of  $B^{*}$ anomaly found in pure UBe$_{13}$.
These findings   lead  to a new channel  to  deepen  its true nature of  the  ground state
 of    U$_{1-x}$Th$_{x}$Be$_{13}$,  clarifying  the origin of the 
 unusual transition inside the SC phase.

\color{black}

\begin{acknowledgments}
We greatly appreciate valuable discussions with  M. Yokoyama, Y. Kono, Y. Haga, H. Amitsuka,  and T. Yanagisawa.
We  also would like to thank K. Mochidzuki and K. Kindo for 
  the use of Magnetic Properties Measurement System (Quantum Design, Inc.) and their supports.
One of us (Y.S.) would like to thank 
 all the  supports from  Institute for Materials Research, Tohoku University  
 \correct{in growing} monocrystalline  U$_{1-x}$Th$_{x}$Be$_{13}$ samples using the joint-research facility at Oarai.
The present work was supported in part by a Grant-in-Aid for Scientific Research on Innovative Areas ``J-Physics'' (15H05883, 15H05884, 15K05882)
 from MEXT, and KAKENHI (15H03682, 15H05745, 15K05158, 16H04006,  26400360, and 17K14328).
\end{acknowledgments}

\bibliography{apssamp}

\begin{thebibliography}{99}




\bibitem{Ott_PRL_1983} H. R. Ott, H. Rudigier, Z. Fisk, and J. L. Smith, Phys. Rev. Lett. {\bf50}, 1595 (1983).

\bibitem{Stewart_PRL_1984} G. R. Stewart, Z. Fisk, J. O. Willis, and J. L. Smith,  Phys. Rev. Lett. {\bf52}, 679 (1984).

\bibitem{URS1} W. Schlabitz, J. Baumann, B. Pollit, U. Rauchschwalbe, H. M. Mayer, U. Ahlheim, and C. D. Bredl, Z. Phys. B {\bf62}, 171 (1986).

\bibitem{URS2}T. T. M. Palstra, A. A. Menovsky, J. van den Berg, A. J. Dirkmaat, P. H. Kes, G. J. Nieuwenhuys, and J. A. Mydosh, Phys. Rev. Lett. {\bf55}, 2727 (1985).


\bibitem{Maple85} M.B. Maple, J.W. Chen, S.E. Lambert, Z. Fisk, J.L. Smith, H.R. Ott, J.S. Brooks, and M.J. Naughton, Phys. Rev. Lett. {\bf 54}, 477 (1985).

\bibitem{MacLaughlin_PRL_1984}
 D. E. MacLaughlin, C. Tien, W. G. Clark, M. D. Lan, Z. Fisk, J. L. Smith, H. R. Ott,
 Phys. Rev. Lett. {\bf 53}, 1833 (1984).

\bibitem{Walti_PRBR_2001} Ch.  W\"{a}lti, E. Felder, H. R. Ott, Z. Fisk, and J. L. Smith, Phys. Rev. B {\bf63}, 100505(R) (2001).


\bibitem{Tien_PRB_1989} C. Tien, and I. M. Jiang, Phys. Rev. B {\bf40}, 229 (1989).

\bibitem{Einzel86} D. Einzel, P.J. Hirschfeld, F. Gross, B.S. Chandrasekhar, K. Andres, H.R. Ott, J. Beuers, Z. Fisk, and J.L. Smith, Phys. Rev. Lett. {\bf 56}, 2513 (1986).

\bibitem{Fomin00} I. A. Fomin, and J.P. Brison, J. Low Temp. Phys. {\bf 119}, 627 (2000).

\bibitem{Ott_PRL_1984} H. R. Ott, H. Rudigier, T. M. Rice, K. Ueda, Z. Fisk, J. L. Smith, 
 Phys. Rev. Lett. {\bf 52}, 1915 (1984).

 
 
\bibitem{YShimizu_PRL_2015} Y. Shimizu, S. Kittaka, T. Sakakibara, Y. Haga, E. Yamamoto, H. Amitsuka, Y. Tsutsumi, and K. Machida, 
 Phys. Rev. Lett. {\bf114}, 147002 (2015).
\color{black}


\bibitem{Smith_Physica_1985}  J. L. Smith, Z. Fisk, J. O. Willis, A. L. Giorgi, R. B. Roof, H. R. Ott, H. Rudigier, and E. Felder, Physica {\bf135B}, 3 (1985). 

\bibitem{Ott_PRBR_1985} H. R. Ott, H. Rudigier, Z. Fisk, and J. L. Smith, Phys. Rev. B {\bf31},  (R)1651 (1985).
\bibitem{Ott_PRBR_1986} H. R. Ott,  H. Rudigier, E. Felder, Z. Fisk, and J. L. Smith, Phys. Rev. B {\bf33}, 126 (1986).


\bibitem{Heffner_PRL_1990} R. H. Heffner, J. L. Smith, J. O. Willis, P. Birrer, C. Baines, F. N. Gygax, B. Hitti, E. Lippelt, H. R. Ott, A. Schenck, E. A. Knetsch, J. A. Mydosh, and D. E. MacLaughlin, Phys. Rev. Lett. {\bf65}, 2816 (1990).



\bibitem{Steglich_RPP_2016} F. Steglich and S. Wirth, Rep. Prog. Phys. {\bf79}, 084502 (2016).

\bibitem{Kenzelmann_RPP_2017} M. Kenzelmann, Rep. Prog. Phys. {\bf80}, 034501 (2017).

\color{black}


\bibitem{Sigrist_RevModPhys_1991} M. Sigrist and K. Ueda, Rev. Mod. Phys. {\bf63}, 239 (1991).;
   M. Sigrist, and T. M. Rice, Phys. Rev. B {\bf39}, 2200 (1989).

\bibitem{Batlogg_PRL_1985} B. Batlogg, D. Bishop, B. Golding, C. M. Varma, Z. Fisk, J. L. Smith, and H. R. Ott, Phys. Rev. Lett. {\bf55}, 1319 (1985).
\bibitem{Machida_PRL_1987} K. Machida, and M. Kato, Phys. Rev. Lett. {\bf58}, 1986 (1987).


\bibitem{Jin_PRBR_1994} D. S. Jin, T. F. Rosenbaum, J. S. Kim, and G. R. Stewart, Phys. Rev. B {\bf49},
 1540 (1994).


\bibitem{Sonier_muSR} J. E. Sonier, R. H. Heffner, D. E. MacLaughlin, G. J. Nieuwenhuys, O. Bernal, R. Movshovich, P. G. Pagliuso, J. Cooley, J. L. Smith, and J. D. Thompson,  Phys. Rev. Lett. {\bf85}, 2821 (2000).
   J. E. Sonier, R. H. Heffner, G. D. Morris,  D. E. MacLaughlin,  O.  O. Bernal,  J. Cooley, J. L. Smith, and J. D. Thompson,
 Physica B  {\bf326}, 414 (2003).


\color{black}



\bibitem{Kromer_PRL_1998} F. Kromer, R. Helfrich, M. Lang, F. Steglich, C. Langhammer, A. Bach, T. Michels, J. S. Kim, and G. R. Stewart, Phys. Rev. Lett. {\bf81}, 4476 (1998).

\bibitem{Kromer_PRB_2000} F. Kromer, M. Lang, N. Oeschler, P. Hinze, C. Langhammer, F. Steglich,  J. S. Kim, and G. R. Stewart, Phys. Rev. B {\bf62}, 12477 (2000).
\bibitem{Ellman_PRBR_1991} B. Ellman, T. F. Rosenbaum, J. S. Kim, and G. R. Stewart, Phys. Rev. B {\bf44}, 12074(R) (1991). 

\bibitem{YShimizu_PRL_2012} Y. Shimizu, Y. Haga, Y. Ikeda, T. Yanagisawa, and H. Amitsuka, Phys. Rev. Lett. {\bf109}, 217001 (2012).



\bibitem{Sakakibara_JJAP_1994} T. Sakakibara, H. Mitamura, T. Tamaya, and H. Amitsuka, Jpn. J. Appl. Phys. {\bf33}, 5067 (1994).
 
\bibitem{Mota2000} A. C. Mota, E. Dumont, J. L. Smith, and Y. Maeno, Physica C {\bf 332}, 272 (2000).

\bibitem{Zieve95} R. J. Zieve, T. F. Rosenbaum, J. S. Kim, G. R. Stewart, and M. Sigrist, Phys. Rev. B {\bf 51}, 12041 (1995).


\bibitem{Hiess_PRB_2002} A. Hiess, R. H. Heffner, J. E. Sonier, G. H. Lander, J. L. Smith, and J. C. Cooley, 
 Phys. Rev. B {\bf66}, 064531 (2002).



\bibitem{Volovik_JETP_Lett_1993}  G. E. Volovik, JETP. Lett. {\bf58}, 469 (1993).
\bibitem{Vekhter_PRBR_1999} I. Vekhter, P. J. Hirschfeld, J. P. Carbotte, and E. J. Nicol, Phys. Rev. B {\bf59}, R 9023 (1999).
\correct{\bibitem{Sakakibara_ROPP_2016} T. Sakakibara, S. Kittaka, and K. Machida,
 Rep. Prog. Phys. {\bf 79}, 094002 (2016). } 

\bibitem{Volovik_JPhysC_21_1988} G. E. Volovik, J. Phys. C: {\bf21}, L221 (1988).; 
  JETP Lett. {\bf 65}, 491 (1997).

\bibitem{Miranovic_PRB_2003} P. Miranovi\'{c}, N. Nakai, M. Ichioka, and K. Machida, Phys. Rev. B {\bf68}, 052501 (2003).
  N. Nakai, P. Miranovi\'{c}, M. Ichioka, and K. Machida, Phys. Rev. B {\bf70}, R 100503 (2004).



\bibitem{Nodal_HeatCapacity}

\correct{In a nodal superconductor},  the field dependence of $C/T$ should exhibit a convex upward curvature at low fields. In the case of line nodes, \correct{in particular, the} heat capacity becomes  $C(H) \propto (H/H_{\mathrm{c2} })^{1/2}$~\cite{Volovik_JETP_Lett_1993, Vekhter_PRBR_1999, Sakakibara_ROPP_2016}, 
  whereas for point nodes, $C(H) \propto \frac{H}{H_{\mathrm{c2} } } \mathrm{ln} \frac{H}{H_{\mathrm{c2} } } $ 
 \cite{Volovik_JPhysC_21_1988},  or $C(H) \propto (H/H_{\mathrm{c2} })^{0.64}$ 
 \cite{Miranovic_PRB_2003}.
For a clean isotropic $s$-wave superconductor, \correct{on the other hand,} $C(H)/T \propto H$  at low fields.





\bibitem{Takegahara_PhysicaB_2000} K. Takegahara and H. Harima, Physica B {\bf281}, 764 (2000).
\bibitem{Maehira_PhysicaB_2002} T. Maehira, A. Higashiya, M. Higuchi, H. Yasuhara, and A. Hasegawa, Physica B {\bf312-313}, 103 (2002).
\bibitem{Volovik_JETP_1985} G. E. Volovik and L. P. Gor'kov, JETP {\bf61}, 843 (1985).
\bibitem{Blount_PRB_1985} E. I. Blount, Phys. Rev. B {\bf32}, 2935 (1985).
\bibitem{Ozaki_Machida_Ohmi_PTP_1985} M. Ozaki, K. Machida, and T. Ohmi, Prog. Theor. Phys. {\bf74}, 221 (1985).




\bibitem{Endnote_Eg_symmetry} One may consider  the even-parity $E_{g}$ state, \textit{i.e.},
  $ l_{1} (k) = \sqrt{3} ( k^2_{x} -  k^2_{y} )$,  $ l_{2} (k) = 2  k^2_{z} -  k^2_{x}  -  k^2_{y}  $, 
 and a linear combination  $ \psi (k) = l_{1} + i l_{2} =   k^2_{x} + \epsilon  k^2_{y} +  \epsilon^2  k^2_{z} $. 
Although $ \psi (k) = l_{1} + i l_{2} $  may  explain the \correct{nodeless gap} as well as spontaneous magnetism in the B phase, 
   both components  ${l}_{1}$ and ${l}_{2} $  have line nodes which cannot explain the absence of nodal quasiparticle excitations in  pure UBe$_{13}$ 
 \cite{YShimizu_PRL_2015}.
 






\bibitem{Endnote_isotropic_A1g}
 Analogously, the case of the combined even-parity  state  is also possible, \textit{e.g.},
  with  fully isotropic $s$-wave $A_{1g}$  state and another gap symmetry, such as a $d$-wave symmetry.





\bibitem{Jin_PRB_1996} D. S. Jin, S. A. Carter, T. F. Rosenbaum, J. S. Kim, and G. R. Stewart, Phys. Rev. B {\bf53}, 8549 (1996).

\bibitem{Yip91PRB} S. K. Yip, T. Li, and P. Kumar, Phys. Rev. B {\bf 43}, 2742 (1991).

\bibitem{Endnote_1stOrderPhaseTransition}  
 \correct{
 An occurrence of a 1st-order phase transition is neccessary when two 2nd-order phase transition lines meet at a bi-critical point.  
To the best of our knowledge, however, there is no evidence for a 1st-order phase transition in U$_{1-x}$Th$_x$Be$_{13}$.}



\bibitem{Machida_JPSJ_1989} 
  K. Machida, M. Ozaki, and T. Ohmi, J. Phys. Soc. Jpn. {\bf58}, 4116 (1989).
\end{thebibliography}


\end{document}